\documentclass[a4paper,11pt, final]{article}
\usepackage[utf8]{inputenc}
\usepackage{jheppub}
\usepackage{amsmath}

\usepackage[inline]{showlabels}

\newcommand*\diff{\mathrm{d}}

\usepackage{xspace}

\newcommand*{\eq}{eq.\@\xspace}
\newcommand*{\eqs}{eqs.\@\xspace}
\newcommand*{\cf}{cf.\@\xspace}

\title{Einstein-Cartan gravity, matter,\newline and scale-invariant generalization}
\author[a]{M. Shaposhnikov,}
\author[a]{A. Shkerin,}
\author[a]{I. Timiryasov,}
\author[a]{S. Zell}

\affiliation[a]{Institue of Physics, Laboratory for Particle Physics and Cosmology,\\
	\'Ecole Polytechnique F\'ed\'erale de Lausanne, CH-1015 Lausanne, Switzerland}

\emailAdd{mikhail.shaposhnikov@epfl.ch}
\emailAdd{andrey.shkerin@epfl.ch}
\emailAdd{inar.timiryasov@epfl.ch}
\emailAdd{sebastian.zell@epfl.ch}

\abstract{

We study gravity coupled to scalar and fermion fields in the Einstein-Cartan framework. We discuss the most general form of the action that contains terms of mass dimension not bigger than four, leaving out only contributions quadratic in curvature. By resolving the theory explicitly for torsion, we arrive at an equivalent metric theory containing additional six-dimensional operators. This lays the groundwork for cosmological studies of the theory. We also perform the same analysis for a no-scale scenario in which the Planck mass is eliminated at the cost of adding an extra scalar degree of freedom. Finally, we outline phenomenological implications of the resulting theories, in particular to inflation and dark matter production.

}

\date{}

\begin{document}
	
\maketitle

\section{Introduction}
\label{sec:intro}

\subsection{Einstein-Cartan gravity vs metric gravity}

Gravity exists in different incarnations. Apart from the original Einstein's metric General Relativity, there is the Einstein-Cartan (EC) formulation \cite{Cartan:1922,Cartan:1923,Cartan:1924,Cartan1925,Einstein1928,Einstein19282,Utiyama:1956sy, sciama1962analogy,Kibble:1961ba}.\footnote
{See, e.g., \cite{Hehl:1976kj, Shapiro:2001rz} for reviews of EC theory.}
In this theory, the role of fundamental fields is played by the tetrad and the spin connection. The metric is derived from the tetrad field and the Christoffel symbols are defined in terms of the metric-compatible connection. In general, the Christoffel symbols are not symmetric, hence the theory contains torsion. EC gravity may have conceptual advantages as compared to the metric formulation. It can be viewed as a gauge theory of the Lorentz group.\footnote{Other gauge theories of gravity are based on gauging the Poincar\'{e} \cite{Hehl:1978yt,Hayashi:1979wj,Sezgin:1979zf} or Weyl \cite{Bregman:1973fv,Charap:1973fi,Kasuya:1975tx} symmetries; see also \cite{blagojevic2013gauge,Krasnov:2017epi} for reviews and \cite{Ivanov:1981wn,Nair:2008yh,Nikiforova:2009qr,Baekler:2010fr,Karananas:2014pxa,Lasenby:2015dba} for further developments.} This puts gravity on the same footing as the fundamental forces of the Standard Model. Besides, since no second derivatives of the metric appear in the Hilbert-Palatini term, no Gibbons-Hawking-York boundary term \cite{York:1972sj, Gibbons:1976ue} is needed for the derivation of equations of motion.  

The metric and EC formulations of pure gravity yield the same theory (see, e.g, \cite{Dadhich:2010xa} and references therein). This changes once matter is introduced. For a scalar field $h$, EC and metric gravity lead to different predictions if $h$ is coupled directly to the Ricci scalar. In this case, EC gravity is equivalent to the Palatini formulation \cite{Palatini1919, Einstein1925} (see also \cite{Ferraris1981}), in which the metric and the Christoffel symbols are viewed as independent variables.

The non-minimal coupling of the scalar field to gravity can be removed by a suitable Weyl transformation of the metric. After that, the non-equivalence of the two versions of gravity manifests itself as the difference in the (non-canonical) kinetic terms of the scalar field. Next, in EC gravity, fermions $\Psi$ source torsion. Therefore, they also cause a difference as compared to the metric formulation of gravity. As in the case of a non-minimally coupled scalar field, one can derive an equivalent metric theory. This is due to the fact that equations of motion for the torsionful part of the connection reduce to a linear constraint, which can be resolved explicitly. As a result, a new four-fermion interaction term appears \cite{Kibble:1961ba, osti_4843429}. As long as we use the same action as in the metric theory, however, this interaction is suppressed by $1/M_P^2$, where $M_P$ is the Planck mass. Thus, it is observationally irrelevant at subplanckian energies.

There is another important difference of the two formulations of gravity: in the EC version, one can use a more general action. The reason is that there are additional operators of mass dimension not bigger than four. After solving for torsion, they lead to new six-dimensional terms in the equivalent metric theory. Three types of operators can appear this way; they are, schematically, of the form $f_1(h)(\partial h)^2$, $f_2(h) (\partial h)\bar{\Psi}\Psi$ and $f_3(h)(\bar{\Psi}\Psi)^2$. It is important to note that in the original EC action, the extra terms come with arbitrary coupling constants. Hence, the mass scale suppressing the six-dimensional operators in the equivalent metric theory is not pre-determined. If this scale is chosen appropriately, the new operators can have phenomenological consequences.

Of course, one could have started from the beginning in a metric theory with additional higher-dimensional operators. In this case, however, a consistent effective field theory approach would dictate that all possible higher-dimensional interactions (consistent with relevant symmetries) are taken into account. In contrast, viewing the EC action as fundamental only leads to a specific subset of higher-dimensional operators. They are the ones that, before solving for torsion, can be obtained without adding any higher-dimensional terms.

In the present work, we study EC gravity coupled to scalars and fermions. We include all terms of mass dimension not bigger than four, except for the terms quadratic in curvature. The presence of such terms would complicate the equation for the torsionful part of the connection, and we leave the study of this more general case for future work.\footnote
{Including the terms quadratic in curvature may be interesting for phenomenology. For example, in Palatini gravity without fermions, the effect of the $R^2$-term has recently been studied in \cite{1810.05536, 1810.10418, 1901.01794, 1902.07876, 1911.11513, 1912.12757, 2002.08324}, mainly in the context of inflation.}
In metric gravity, the only possible terms would be the Ricci scalar $R$ and a non-minimal coupling term $h^2 R$.\footnote
{We do not consider the cosmological constant term since it is independent of the connection and, therefore, insensitive to the difference between metric and EC gravity.}
In contrast, the following additional terms appear in the EC formulation:
\begin{itemize}
	\item Since the Riemann tensor loses its antisymmetry property, contracting indices in the Ricci tensor is no longer the only way to form a scalar: another invariant appears which is called the Holst term \cite{Hojman:1980kv, Nelson:1980ph, Castellani:1991et, Holst:1995pc}.
	\item As for the Ricci scalar, the scalar field can be coupled directly to the Holst term.
	\item Moreover, there is the topological invariant corresponding to the Nieh-Yan class \cite{Nieh:1981ww}. By itself, it contributes to the boundary term. However, when coupled to the scalar field, it gives a nontrivial contribution to equations of motion.
	\item Finally, it is possible to extend the kinetic term of fermions. Whereas it is unique in metric gravity, one can write two additional nontrivial terms in the presence of torsion \cite{hep-th/0507253, Alexandrov:2008iy,1104.2432, 1212.0585}. 
\end{itemize}
To summarize, we get one additional coupling constant due to gravity and two extra couplings per scalar and fermion fields. In this paper, we derive and analyze the metric theory corresponding to this general EC theory. Our study generalizes the works \cite{Nelson:1980ph, Castellani:1991et, gr-qc/0505081, hep-th/0507253,0807.2652, 0811.1998, 0902.0957, 0902.2764, 1104.2432, 1212.0585} in which the metric theory is derived for different subsets of the above-mentioned terms.

\subsection{Motivation}

One motivation of our study is the general question about what the fundamental theory of gravity might look like. For example, the canonical formulation of General Relativity by Ashtekar and Barbero \cite{Ashtekar:1986yd,Barbero:1994ap}, which includes the Holst term, is a starting point for quantization in Loop Quantum Gravity (see \cite{Thiemann:2007zz} and references therein). Furthermore, we also have in mind various phenomenological implications. Applications of EC gravity to particle physics and cosmology have been investigated, e.g., in \cite{hep-th/0507253, 0807.2652, 0811.1998,1104.2432, 1201.4226, 1201.5423, 1212.0585, Khriplovich:2013tqa}. The list of addressed topics includes observational signatures of parity violation in quantum gravity \cite{hep-th/0507253}, possible signatures of four-fermion interaction terms in experiment \cite{hep-th/0507253,1104.2432} and their possible relevance for cosmological evolution \cite{1201.5423,1212.0585, Khriplovich:2013tqa}, inflation driven by the Holst term \cite{0807.2652, 0811.1998}. In the latter case, the coefficient of the Holst term (the Barbero-Immirzi parameter) is promoted to a dynamical field \cite{0807.2652, 0811.1998, 0902.0957} which makes it resemble the non-minimally coupled scalar field discussed above. 

Our first phenomenological interest in EC gravity is also related to inflation \cite{secondPaper}. However, we adopt a different point of view regarding the nature of the dynamical field coupled to the Holst term. Namely, we associate this field with the Higgs field of the Standard Model. This is motivated by the well-known fact that, once the Higgs field is coupled directly to the Ricci scalar, it can serve as an inflaton \cite{Bezrukov:2007ep}. The resulting model of Higgs inflation has been studied both in the metric \cite{Bezrukov:2007ep} and Palatini \cite{Bauer:2008zj} formulations of gravity. Because of the non-minimal coupling, the two scenarios lead to different predictions, although both are fully compatible with the current CMB observations.\footnote
{An advantage of Palatini Higgs inflation is that the connection between inflationary physics and parameters of the Standard Model as measured in collider experiments may be more robust than in the metric theory \cite{Shaposhnikov:2020fdv}.}
As stated, Palatini gravity is equivalent to the EC theory in the absence of fermions and additional terms. As soon as they are included, the immediate question is to what extent they change inflation. By deriving the dimension-six interactions of the equivalent metric theory, the present work forms the basis for this study that is continued in \cite{secondPaper}.

The EC framework looks particularly attractive in the context of our proposal for combining the Standard Model with gravity \cite{Shaposhnikov:2020geh}. There, the important ingredients are the non-minimal coupling of the Higgs field to gravity and the Palatini formulation of gravity. Moreover, we considered a scenario in which the Standard Model is classically scale-invariant, i.e., the tree-level Higgs mass is zero, and no new degrees of freedom exist anywhere above the Electroweak scale. Then, not only the model incorporates Palatini Higgs inflation, but it can also host the non-perturbative gravitational mechanism of Electroweak symmetry breaking suggested in \cite{Shaposhnikov:2018xkv} and studied further in \cite{Shaposhnikov:2018jag,Shkerin:2019mmu,Karananas:2020qkp}. Once all particles of the Standard Model are included, however, the Palatini formulation of gravity is no longer appropriate. The reason is that the coupling of fermions to gravity is realized by the spin connection. Therefore, one should use the spin connection as dynamical variables instead of Christoffel symbols. In this way, the Palatini formulation of gravity naturally generalizes to EC theory, providing strong motivation to study EC gravity and its phenomenological consequences.

Another motivation to study the EC theory is related to four-fermion interaction terms appearing in the action after solving for torsion. The terms come with arbitrary couplings and choosing them appropriately may have phenomenological consequences. In the course of cosmological evolution, they can lead to the production of particles that otherwise only interact very weakly. In particular, we will show in \cite{thirdPaper} that the four-fermion interaction can provide a dominant channel for singlet (with respect to the Standard Model gauge group) fermion production. In the setting of the Neutrino Minimal Standard Model ($\nu$MSM) \cite{Asaka:2005an,Asaka:2005pn}, whose particle content is extended compared to that of the Standard Model only by three Majorana neutrinos with masses below the Electroweak scale, this means that sterile neutrinos can be produced in this way and can account for all of dark matter even if their Yukawa couplings to leptons are equal to zero.

\subsection{Einstein-Cartan gravity and global scale symmetry}

All studies of EC gravity up to date were carried out while keeping explicitly $M_P$ (and the cosmological constant) as dimensionful parameters. Meanwhile, one can also consider a scenario in which gravity enjoys a global scale symmetry. In order to achieve this, one adds a new degree of freedom -- a scalar dilaton -- to the theory. Then the Planck scale arises due to a spontaneous breaking of the scale symmetry by a vacuum expectation value of the dilaton. Scale-invariant extensions of the Standard Model and General Relativity were studied extensively in the literature. This was mostly done in the metric formulation \cite{Wetterich:1987fm,Shaposhnikov:2008xb,GarciaBellido:2011de,Bezrukov:2012hx,Ferreira:2016wem,Ferreira:2016kxi,Ferreira:2018itt}\footnote{For a recent review of fundamental scale invariance, see \cite{Wetterich:2020cxq}.} but more recently also in the Palatini one \cite{2004.00039}. They exhibit promising applications both to early- and late-time cosmology.

In this paper, we lay the groundwork for extending these studies to EC gravity. Concretely, we consider a scenario in which gravity in the EC formulation is coupled non-minimally to two scalar fields. One of them -- the dilaton -- is responsible for generating the Planck mass at low energies. Another one is associated with the Higgs field degree of freedom once the gravitational action is combined with the Standard Model. The two scalar fields are provided with the interaction potential that gives rise to the Higgs mass and the cosmological constant. We show that the resulting theory promises to inherit vivid phenomenological implications from its metric counterparts such as the Higgs-Dilaton model \cite{GarciaBellido:2011de,Bezrukov:2012hx}.

The paper is organized as follows. In section \ref{sec:EC-Higgs}, we introduce the theory with one scalar field and fermions and discuss the various contributions to its action. Next, we solve the theory for torsion to obtain an effective action in the metric formulation. We study the consistency of the resulting theory and discuss parameter limits that reproduce known results of both metric and Palatini gravity. In section \ref{sec:EC-Higgs-Dilaton}, we repeat the analysis for a classically scale-invariant theory of EC gravity, two scalar fields and free fermions. Again, we obtain an effective action and discuss the consistency of the resulting theory. Finally, in section \ref{sec:conclusion} we discuss phenomenological implications of the theory and outline directions for future research.

\textbf{Conventions}. We work in natural units $\hbar=c=1$ and use the metric signature $(-1,+1,+1,+1)$. The indices $I,J,\ldots$ run from 0 to 3. The antisymmetric tensor is defined by $\epsilon_{0123}=1$. For compactness, we omit the wedge product sign. The matrix $\gamma^5$ is defined as $\gamma^5=\frac{1}{i}\gamma^0\gamma^1\gamma^2\gamma^3$.

\section{Einstein-Cartan gravity with one scalar field}
\label{sec:EC-Higgs}

\subsection{General action}
We consider four-dimensional local Lorentz-invariant, general covariant theory of EC gravity extended by one scalar degree of freedom as well as fermions. As far as phenomenological implications are concerned, the scalar degree of freedom will be associated with the Higgs field in the unitary gauge, but this identification is inessential for our subsequent analysis. We assume that the action  only contains leading bulk terms of mass dimension not greater than four and polynomial in fields and their derivatives. In the present analysis, we leave out terms quadratic in the curvature. Aside from this omission, the most general gravitational action of the theory reads as follows:
\begin{equation}\label{S_gr}
\begin{split}
    S_{gr}= & \frac{1}{4}\int (M_P^2+\xi_hh^2)\epsilon_{IJKL}e^Ie^JF^{KL} + \frac{1}{2\bar{\gamma}}\int (M_P^2+\xi_\gamma h^2)e^Ie^JF_{IJ} + \frac{1}{2}\int \xi_\eta h^2\diff (e^Ie^JC_{IJ}) \;,
\end{split}
\end{equation}
where $M_P$ is the reduced Planck mass. Moreover, $e^I$ is a tetrad one-form, $F^{IJ}=\diff \omega^{IJ}+\omega^I_K\omega^{KJ}$ represents the curvature two-form with $\omega^{IJ}$ the Lorentz connection one-form, and $C_{IJ}$ is the contorsion one-form related to the Lorentz connection as
\begin{equation}
    \omega^{IJ}=\mathring{\omega}^{IJ}+C^{IJ} \;, ~~~ \mathring{D}e^I=0 \;.
\end{equation}
We assume that $C_{IJ}$ is antisymmetric which implies zero non-metricity. In the first term in the r.h.s. of \eq (\ref{S_gr}) we recognize the Hilbert-Palatini action to which the Higgs field is coupled directly with the coupling $\xi_h$. The second term represents the Holst action \cite{Hojman:1980kv, Nelson:1980ph, Castellani:1991et, Holst:1995pc} with the coefficient $\bar{\gamma}$ called the Barbero-Immirzi parameter \cite{Immirzi:1996dr,Immirzi:1996di}.\footnote{Note that the Holst term is a pseudoscalar under spacetime reflections. Hence, if one wants the theory to be parity-invariant, one has to treat $\bar{\gamma}$ as a pseudoscalar; see, e.g., \cite{0811.1998}.} Moreover, we have a second non-minimal coupling $\xi_\gamma$.
Finally, the third term contains the Nieh-Yan invariant \cite{Nieh:1981ww}. In a theory without non-minimal couplings, it represents a boundary term. In our theory, however, it leads to a non-trivial contribution because of the third non-minimal coupling $\xi_\eta$. We note that the remaining four-dimensional invariants,
\begin{equation}
    \int \epsilon_{IJKL}F^{IJ}F^{KL} \;, ~~~ \int F^{IJ}F_{IJ} \;,
\end{equation}
representing the Euler and Pontryagin topological classes correspondingly, are operators of dimension four. Hence, they are not coupled to the scalar fields and do not participate in classical dynamics. In total, in the action (\ref{S_gr}) we have four dimensionless couplings:
\begin{equation}\label{params}
    \xi_h \;, ~~ \xi_\gamma \;, ~~ \xi_\eta  ~~ \rm{and} ~~ \bar{\gamma} \;.
\end{equation}

To make the scalar field dynamical, we supplement the action (\ref{S_gr}) with the kinetic term for $h$. We also add the potential term $U$ so that the scalar field action takes the form
\begin{equation}\label{S_s}
    S_s=\int \frac{\epsilon_{IJKL}e^Ie^Je^Ke^L}{24} \left( -\frac{1}{2} (\partial_N h)^2- U\right) \;.
\end{equation}
When $h$ is associated with the Higgs field, we get $U=\frac{\lambda}{4}(h^2-v^2)^2$. Here the parameter $v$ becomes the Electroweak vacuum expectation value and $\lambda$ represents the Higgs self-coupling constant.

Finally, we turn to the fermionic part of the action. For simplicity, in what follows we restrict ourselves to a single fermion generation. One can write the following general fermion kinetic term \cite{hep-th/0507253, 1104.2432, 1212.0585}:
\begin{equation}\label{S_f}
    S_f=\frac{i}{12}\int \epsilon_{IJKL}e^Ie^Je^K\left(\bar{\Psi}(1-i\alpha-i\beta\gamma^5)\gamma^LD\Psi-\overline{D\Psi}(1+i\alpha+i\beta\gamma^5)\gamma^L\Psi\right) \;,
\end{equation}
where $D\Psi=d\Psi+\frac{1}{8}\omega_{IJ}[\gamma^I,\gamma^J]\Psi$ and $\gamma^I$ are the gamma matrices. The real parameters
\begin{equation}\label{params2}
    \alpha \;, ~~ \beta
\end{equation}
are non-minimal fermion couplings. The corresponding terms in eq.~(\ref{S_f}) vanish in the case of zero torsion, but in the general case, they contribute to the interactions between the fermionic currents in the effective metric theory.

The total action we are interested in takes the form
\begin{equation}\label{S_tot}
    S=S_{gr}+S_s+S_f \;.
\end{equation}
Note that if the Higgs vacuum expectation value $v$ is put to zero, then the total action is manifestly invariant under the scale transformations 
\begin{equation}\label{ScaleTr}
    h\mapsto ph  \;, ~~ e^I\mapsto\frac{e^I}{p} \;, ~~ \omega^{IJ}\mapsto\omega^{IJ} \;, ~~ \Psi\mapsto p^{3/2}\Psi
\end{equation}
with constant $p$. As discussed in the introduction, in a scenario with no Electroweak symmetry breaking at tree level, it is possible to generate the observed value of $v$ via a non-perturbative gravitational mechanism \cite{Shaposhnikov:2020geh}.

\subsection{Solving for torsion}

Our immediate goal is to bring the theory (\ref{S_tot}) to the form suitable for phenomenological analysis. To this end, one proceeds in two steps. First, one gets rid of the non-minimal coupling to the Ricci scalar using a conformal transformation. Secondly, the resulting theory admits an explicit solution for the contorsion form $C_{IJ}$, and one can use it to rewrite the theory effectively in metric gravity. Upon such rewriting, multiple higher-order interaction terms between the scalar field and fermions appear. 

The first step is achieved by performing the transformation (\cf \eq (\ref{ScaleTr})):
\begin{equation}\label{ConfTr}
    e^I\mapsto\frac{e^I}{\Omega} \;, ~~ \omega^{IJ}\mapsto \omega^{IJ} \;, ~~ \Psi\mapsto \Omega^{3/2}\Psi, ~~ \Omega^2=1+\frac{\xi_hh^2}{M_P^2} \;.
\end{equation}
The gravitational action becomes
\begin{equation}\label{S_gr_tr}
    S_{gr}\mapsto\frac{M_P^2}{4}\int \left\lbrace \epsilon_{IJKL}e^Ie^JF^{KL}+2\gamma e^Ie^JF_{IJ}+2\eta \diff \left(\frac{e^Ie^JC_{IJ}}{\Omega^2}\right) \right\rbrace \;,
\end{equation}
where 
\begin{equation}\label{GammaEta}
\gamma(h)=\frac{1+\frac{\xi_{\gamma} h^2}{M_P^2}}{\bar{\gamma}\Omega^2} \,, ~~~ \eta(h)=\frac{\xi_\eta h^2}{M_P^2} \;.
\end{equation}
One can think of $\gamma(h)$ as a field-dependent Barbero-Immirzi parameter. The action of the scalar fields becomes
\begin{equation}\label{S_s_tr}
     S_s\mapsto\int \frac{\epsilon_{IJKL}e^Ie^Je^Ke^L}{24} \left( -\frac{1}{2} \frac{(\partial_N h)^2}{\Omega^2}- \frac{U}{\Omega^4}\right) \;.
\end{equation}
Finally, the fermion action changes as follows:
\begin{equation}\label{S_f_tr}
S_f \mapsto S_f + \frac{1}{8}\int \frac{\epsilon_{I J K L}e^Ie^Je^K\diff\Omega^2}{\Omega^2}(\alpha V^L + \beta A^L) \;,
\end{equation}
where we have introduced the vector and axial fermion currents,
\begin{equation}
    V^I=\bar{\Psi}\gamma^I\Psi \;, ~~~ A^I=\bar{\Psi}\gamma^5\gamma^I\Psi \;.
\end{equation}
Note that in the absence of non-minimal fermion couplings the action $S_f$ is invariant under the conformal transformation as it should.

To find the equation of motion for $C^{IJ}$, we vary the action with respect to the spin-connection. The solution to the equation of motion is then substituted back, yielding the effective metric theory. We follow the procedure carried out, e.g., in \cite{hep-th/0507253,0811.1998,Dadhich:2010xa}. The variation of the gravitational action gives
\begin{equation}
    \frac{\delta S_{gr}}{\delta\omega^{IJ}}=-\frac{M_P^2}{4}\left( D\left( \epsilon^{IJKL}e_Ke_L+2\gamma e^Ie^J \right) + \frac{2\diff\eta}{\Omega^2}e^Ie^J \right) \;.
\end{equation}
Variation of the fermionic part yields
\begin{equation}
\begin{split}
    \frac{\delta S_f}{\delta\omega^{IJ}} & =\frac{i}{12\cdot 8}\epsilon_{KLMN}e^Ke^Le^M\bar{\Psi}\left( \lbrace \gamma^N,[\gamma^I,\gamma^J]\rbrace-i(\alpha+\beta\gamma^5)[\gamma^N,[\gamma^I,\gamma^J]] \right)\Psi \\
    & =\frac{1}{24}\epsilon_{KLMN}e^Ke^Le^M\left( \epsilon^{NIJP}A_P+2\delta^{N[I}\left( \alpha V^{J]}+\beta A^{J]} \right) \right) \;,
\end{split}
\end{equation}
The brackets $\{\}$ and $[]$ denote symmetrization and antisymmetrization, respectively. The equation of motion for $C^{IJ}$ reads as follows:\footnote
{It is important to note that \eq (\ref{EoM_for_C}) is algebraic with respect to the torsionful part $C_{IJ}$ of the connection, hence the latter does not give rise to new degrees of freedom.}
\begin{equation}\label{EoM_for_C}
\begin{split}
-2 C^{[I}_{\ K}B^{J]K} + &  \left( \frac{\diff \eta}{\Omega^2}+\diff \gamma\right)e^I e^J \\
& = \frac{1}{M_P^2} \left(\frac{1}{2} e^I e^J e^P A_P+ \frac{1}{6} \epsilon_{KLMN} e^K e^L e^M \delta^{N[I}\left(\alpha V^{J]} + \beta A^{J]}\right)\right) \;,
\end{split}
\end{equation}
where
\begin{equation}
B^{KJ} = \frac{1}{2} \epsilon^{KJLM} e_L e_M + \gamma e^K e^J \;.
\end{equation}
The solution to this equation is given by
\begin{equation}\label{C_IJ}
\begin{split}
C^{IJ} & = - \frac{1}{2(\gamma^2+1)} 
\left( \epsilon^{IJKL} e_K \left( \frac{\partial_L \eta}{\Omega^2} +\partial_L\gamma \right) - 2\gamma e^{[I}\left( \frac{\partial^{J]}\eta}{\Omega^2}+\partial^{J]}\gamma \right)  \right) \\
& + \frac{1}{4M_P^2(\gamma^2+1)} \left( \epsilon^{IJKL} e_K \left(A_L + \gamma (\alpha V_L + \beta A_L)\right) +
2e^{[I} \left(\alpha V^{J]} + (\beta-\gamma) A^{J]}\right) \right) \;.
\end{split}
\end{equation}
It remains to plug this into eqs.~(\ref{S_gr_tr}), (\ref{S_s_tr}), and (\ref{S_f_tr}). The result in component notation is
\begin{subequations}\label{S_eff}
\begin{align}
    S^{\rm eff} & = \int \diff^4x\sqrt{-g}\left\lbrace \frac{M_P^2}{2}\mathring{R}+\frac{i}{2}\overline{\Psi}\gamma^\mu\mathring{D}_\mu\Psi-\frac{i}{2}\overline{\mathring{D}_\mu\Psi}\gamma^\mu\Psi \right\rbrace \\
    & - \int \diff^4x\sqrt{-g} \left\lbrace \frac{1}{2\Omega^2}(\partial_\mu h)^2+\frac{U}{\Omega^4} \right\rbrace \\
    & - \int \diff^4x\sqrt{-g}\frac{3M_P^2}{4(\gamma^2+1)}\left( \frac{\partial_\mu\eta}{\Omega^2}+\partial_\mu\gamma \right)^2\\
    & + \int \diff^4x\sqrt{-g}\frac{3\alpha}{4}\left( \frac{\partial_\mu\Omega^2}{\Omega^2}+\frac{\gamma}{\gamma^2+1}\left( \frac{\partial_\mu\eta}{\Omega^2}+\partial_\mu\gamma \right) \right) V^\mu \\
    & + \int \diff^4x\sqrt{-g}\frac{3}{4}\left( \beta\frac{\partial_\mu\Omega^2}{\Omega^2}+\frac{1+\gamma\beta}{\gamma^2+1}\left( \frac{\partial_\mu\eta}{\Omega^2}+\partial_\mu\gamma \right) \right) A^\mu \\
    & -\int \diff^4x\sqrt{-g}\frac{3}{16M_P^2(\gamma^2+1)}\left( \left( 1+2\gamma\beta-\beta^2 \right)A_\mu^2+2\alpha\left(\gamma-\beta\right)A_\mu V^\mu -\alpha^2 V_\mu^2 \right) \;,
\end{align}
\end{subequations}
where $\gamma$ and $\eta$ are given in \eq (\ref{GammaEta}). Thus, we have obtained an effective action in metric gravity.  The first line in \eq (\ref{S_eff}) contains the standard Einstein-Hilbert term and the fermion kinetic term in the metric formulation of gravity. The second line represents the original action for the scalar field (\ref{S_s}) which underwent the conformal transformation (\ref{ConfTr}). The next four lines represent the various additional scalar-scalar, scalar-fermion and fermion-fermion interaction terms. The scalar-scalar interaction contributes to the kinetic term of $h$, as discussed later. We observe that the Holst and the Nieh-Yan operators contribute in a similar fashion to the effective action (see, e.g., \cite{0902.2764}). Next, we note that in the absence of the non-minimal fermion couplings, the torsion induces only the axial current interaction $\propto A_\mu A^\mu$. In the general case, the vector-vector and the axial-vector couplings are also present; the latter is not invariant under the parity transformation. The three couplings are determined by the three independent parameters. Hence, on the theory side there are no restrictions on the values of the couplings.\footnote{A physical interpretation of the Barbero-Immirzi parameter within Loop Quantum Gravity determines it to be $\gamma\approx 0.274$ \cite{Khriplovich:2001qv}, but here we do not rely on any particular theory possibly complementing the model (\ref{S_tot}) at high energies.} 

\subsection{Consistency and known limits}
The action (\ref{S_eff}) generalizes several known results. When $\beta=\xi_\eta=0$ (and, hence, $\eta=0$), we recover the results of \cite{0811.1998}. For $\bar{\gamma}\to\infty$, $\xi_\eta=0$ (that is, $\gamma=\eta=0$), we reproduce the findings of \cite{1212.0585}. Finally, the case $\alpha=\beta=\gamma=0$ yields the result of \cite{0902.2764}.

We study the scalar sector of the theory (\ref{S_eff}) in \cite{secondPaper}. For reader's convenience, we give a short outlook to some of the results. The kinetic term for $h$ can be written as $-\frac{1}{2}g^{\mu\nu}K(h)\partial_\mu h\partial_\nu h$ where
\begin{equation} \label{Kgeneral}
    K(h)=\frac{1}{\Omega^2}+\frac{6h^2}{\Omega^4M_P^2}\frac{\left( \frac{\xi_\gamma-\xi_h}{\bar{\gamma}}+\xi_\eta\Omega^2\right)^2}{\Omega^4+\frac{1}{\bar{\gamma}^2}\left( 1+\frac{\xi_\gamma h^2}{M_P^2}\right)^2} \;,
\end{equation}
and $\Omega$ is defined in \eq (\ref{ConfTr}). A sufficient condition for the consistency of the theory is that all the couplings (\ref{params}) are non-negative since in this case the function $K(h)$ is positive everywhere. However, more general parameter choices may also be possible.
By choosing the couplings (\ref{params}) appropriately, one can reproduce the known models of scalar-tensor gravity. Consider first the limit of vanishing Holst term, $\bar{\gamma}\to\infty$. The scalar field kinetic term becomes
 \begin{equation}
 \left.K(h)\right\vert_{\bar{\gamma}=\infty}=\frac{1}{\Omega^2}+\frac{6\xi_\eta^2h^2}{\Omega^4M_P^2} \;.
 \end{equation}
Taking further the limit $\xi_\eta=0$, we recover the model of Palatini gravity with the non-minimally coupled scalar field \cite{Bauer:2008zj}. On the other hand, at $\xi_\eta=\xi_h$ the metric version of the same model is reproduced. By varying $\xi_\eta$ from $0$ to $\xi_h$, one continuously deforms the Palatini formulation of the model into its metric formulation.

Consider now the limit $\bar{\gamma}=\xi_\gamma=0$. Note that when the Barbero-Immirzi parameter vanishes, the Holst term becomes singular. This can be amended by introducing an auxiliary one-form field $B^I$ (see, e.g., \cite{hep-th/0507253}). We get, up to a boundary term,
\begin{equation}
    \begin{split}
        \frac{1}{2\bar{\gamma}}\int (M_P^2+\xi_\gamma h^2)e^Ie^JF_{IJ}=-2M_P^2\int (B^IT_I+\bar{\gamma}B^IB_I ) + \frac{1}{2\bar{\gamma}}\int \xi_\gamma h^2e^Ie^JF_{IJ} \;,
    \end{split}
\end{equation}
where $T^I=D e^I$. When $\xi_\gamma=0$, the equation of motion for $B^I$ in the limit $\bar{\gamma}\to 0$ implies $T^I=0$, and the metric formulation of the theory is restored. One can see this explicitly by evaluating $K(h)$ in this limit:
\begin{equation}
    \left.K(h)\right\vert_{\bar{\gamma}=\xi_\gamma=0}=\frac{1}{\Omega^2}+\frac{6\xi_h^2h^2}{\Omega^4M_P^2} \;.
\end{equation}
Thus, in the limit $\bar{\gamma}=\xi_\gamma=0$, the model of metric gravity with the non-minimally coupled scalar field is reproduced.

\section{Scale-invariant model of Einstein-Cartan gravity}
\label{sec:EC-Higgs-Dilaton}

Let us now discuss the no-scale scenario \cite{Shaposhnikov:2008xb,GarciaBellido:2011de,Bezrukov:2012hx, 2004.00039} in which the theory possesses no dimensionful parameters at the classical level.\footnote{ For the discussion of how the scale symmetry can be preserved at quantum level, see \cite{Shaposhnikov:2008xb,GarciaBellido:2011de,Bezrukov:2012hx}.} This is achieved by introducing a new scalar field -- dilaton $\chi$ -- whose vacuum expectation value gives rise to the Planck mass and whose coupling to the Higgs field gives rise to the tree-level Higgs mass (if any). The gravitational action takes the form (\cf \eq (\ref{S_gr}))
\begin{equation}\label{S_gr_d}
\begin{split}
    S_{gr}= & \frac{1}{4}\int (\zeta_\chi\chi^2+\xi_hh^2)\epsilon_{IJKL}e^Ie^JF^{KL} + \frac{1}{2\bar{\gamma}}\int (\zeta_\chi\chi^2+\xi_\gamma h^2)e^Ie^JF_{IJ} \\
    & + \frac{1}{2}\int (\zeta_\eta\chi^2+\xi_\eta h^2)\diff(e^Ie^JC_{IJ}) \;,
\end{split}
\end{equation}
and the scalar field part of the theory reads as follows:
\begin{equation}\label{S_s_d}
S_s=\int \frac{\epsilon_{IJKL}e^Ie^Je^Ke^L}{24} \left( -\frac{1}{2} (\partial_N h)^2-\frac{1}{2} (\partial_N \chi)^2 - U(\chi,h) \right)\;,
\end{equation}
where the scale-invariant scalar field potential is given by
\begin{equation}
    U(\chi,h)=\frac{\lambda}{4}\left(h^2-\frac{a}{\lambda}\chi^2\right)^2+b\chi^4 \;.
\end{equation}
Applied to phenomenology, the parameter $a$ leads to the tree-level mass of the Higgs field and $b$ is responsible for the cosmological constant. Finally, the free fermion action is still given by \eq (\ref{S_f}). Apart from the six parameters (\ref{params}), (\ref{params2}), the scale-invariant theory possesses two additional non-minimal couplings 
\begin{equation}\label{params3}
    \zeta_\chi ~~ \rm{and} ~~ \zeta_\eta \;.
\end{equation}

To find the expression for $C_{IJ}$, we proceed as in section \ref{sec:EC-Higgs}. The only difference is in the definitions of the functions $\gamma$ and $\eta$ which are now given by
\begin{equation}\label{Gamma_Eta_d}
    \gamma(\chi,h)=\frac{\zeta_\chi\chi^2+\xi_\gamma h^2 }{ \bar{\gamma}M_P^2\Omega^2} \;, ~~ \eta(\chi,h)=\frac{\zeta_\eta\chi^2+\xi_\eta h^2}{M_P^2} \;,
\end{equation}
and in the form of the conformal transformation used to get rid of the non-minimal coupling in the Hilbert-Palatini term:
\begin{equation}\label{Omega_d}
    \Omega^2=\frac{\zeta_\chi\chi^2+\xi_hh^2}{M_P^2} \;.
\end{equation}
The contorsion $C_{IJ}$ is still given by \eq (\ref{C_IJ}), where now $\gamma$, $\eta$ and $\Omega$ are defined by \eqs (\ref{Gamma_Eta_d}) and (\ref{Omega_d}). Substituting the solution back into the action results in the effective theory (\ref{S_eff}) with the second line replaced by 
\begin{equation} \label{S_eff_d}
    -\int \diff^4x\sqrt{-g}\left\lbrace \frac{1}{2\Omega^2}(\partial_\mu h)^2+\frac{1}{2\Omega^2}(\partial_\mu\chi)^2+\frac{U(\chi,h)}{\Omega^4} \right\rbrace \;.
\end{equation}

The scale symmetry of the theory can be broken spontaneously by the dilaton field. If we neglect the cosmological constant, i.e., set $b = 0$, the theory admits the classical ground state with $\chi=\bar{\chi}$ and $h^2=\bar{\chi}^2 a/\lambda$. Then the classical vacuum expectation value $\bar{\chi}$ is related to the Planck mass as follows:
\begin{equation}
    M_P^2=\zeta_\chi\bar{\chi}^2 \;.
\end{equation}
A ground state also exists for non-vanishing $b$ \cite{GarciaBellido:2011de}. In this case, spontaneous symmetry breaking simultaneously generates the Planck scale, the vacuum expectation value of the Higgs field and the cosmological constant.

The kinetic term of the fields $\chi$ and $h$ is of the form
\begin{equation}
    -\frac{1}{2}g^{\mu\nu}K_{ab}(\varphi^1,\varphi^2)\partial_\mu\varphi^a\partial_\nu\varphi^b \;,
\end{equation}
where $a,b=1,2$ and $\varphi^1\equiv\chi$, $\varphi^2\equiv h$. It is straightforward to compute the eigenvalues of the matrix $K_{ab}$. They are given by
\begin{equation}
    \lambda_1(\chi,h)=\frac{M_P^2}{\zeta_\chi\chi^2+\xi_hh^2} \;,
\end{equation}
\begin{equation}
\begin{split}
 \lambda_2(\chi,h) & =  M_P^2\left(\zeta_\chi\chi^2+\xi_hh^2 \right)^{-2} \\
& \times\left(\chi ^4 \left(\bar{\gamma
   }^2+1\right) \zeta_\chi^2+h^4 \left(\bar{\gamma }^2 \xi_h^2+\xi_{\gamma}^2\right)+2 h^2 \chi^2 \zeta_{\chi } \left(\bar{\gamma }^2 \xi_h+\xi_{\gamma}\right)\right)^{-1} \\
& \times\bigg( \chi^6 \zeta _{\chi }^2 \left(\bar{\gamma }^2 \left(6 \zeta _{\eta }^2+\zeta _{\chi
   }\right)+\zeta _{\chi }\right)+h^6 \xi _h \left(\bar{\gamma }^2 \xi _h \left(6 \xi_\eta^2+\xi _h\right)+\xi _{\gamma }^2\right) \\  
 & ~~~~~ + h^4 \chi ^2 \left(3 \bar{\gamma }^2 \xi _h
   \left(2 \zeta _{\eta}^2 \xi _h+\zeta _{\chi } \left(4 \xi_\eta^2+\xi
   _h\right)\right)+12 \bar{\gamma } \xi _h \zeta _{\chi } \left(\xi _h-\xi _{\gamma
   }\right) \left(\zeta _{\eta}-\xi_\eta\right) \right.\\
   & ~~~~~ \left. +\zeta _{\chi } \left(\xi _{\gamma }^2
   \left(6 \zeta _{\chi }+1\right)+2 \xi _{\gamma } \xi _h \left(1-6 \zeta _{\chi }\right)+6
   \xi _h^2 \zeta _{\chi }\right)\right)\\
   & ~~~~~ +h^2 \chi ^4 \zeta _{\chi } \left(3 \bar{\gamma }^2
   \left(4 \zeta _{\eta}^2 \xi _h+\zeta_{\chi } \left(2 \xi_\eta^2+\xi
   _h\right)\right)+12 \bar{\gamma } \zeta _{\chi } \left(\xi _h-\xi _{\gamma }\right)
   \left(\zeta _{\eta }-\xi_\eta\right) \right.\\
   & ~~~~~ \left. +\zeta_{\chi } \left(2 \xi_{\gamma } \left(3 \xi_{\gamma }+1\right)-12 \xi_{\gamma } \xi_h+6 \xi_h^2+\xi_h\right)\right) \bigg) \;.
\end{split}
\end{equation}
A sufficient condition for the consistency of the theory is that both eigenvalues are positive, which is achieved if all couplings are non-negative. In this case, the kinetic term of the scalar fields is positive-definite. But as for \eq \eqref{Kgeneral}, more general choices may be admissible.

Again, it is interesting to consider particular regions in the parameter space of the theory. Note that contrary to the case considered in section \ref{sec:EC-Higgs}, the metric formulation of the theory is not recovered in the limit $\bar{\gamma}\to 0$. On the other hand, in the limit $\bar{\gamma} \rightarrow \infty$ of vanishing Holst term we obtain
\begin{equation}
    \left.K\right\vert_{\bar{\gamma}\to\infty}=\frac{1}{M_P^2\Omega^4}\left( \begin{array}{cc}
        (\zeta_\chi+6\zeta_\eta^2)\chi^2+\xi_hh^2 & 6\zeta_\eta\xi_\eta \chi h  \\
        6\zeta_\eta\xi_\eta \chi h & \zeta_\chi\chi^2+(\xi_h+6\xi_\eta^2)h^2
    \end{array} \right) \;.
\end{equation}
For $\zeta_\eta=\xi_\eta=0$, we recover the Palatini formulation and the choice $\zeta_\eta=\zeta_\chi$, $\xi_\eta=\xi_h$ yields the metric version of the scale-invariant model with two scalar fields \cite{Shaposhnikov:2008xb, 2004.00039}.

\section{Discussion and outlook}
\label{sec:conclusion}

There is no doubt that classical gravity is successfully described by General Relativity. However, this still leaves open a question about which formulation of General Relativity one should use. An important alternative to the commonly-used metric formulation is the Einstein-Cartan (EC) version, on which we focused in the present work. In the absence of matter, this question is of purely aesthetical nature since both theories are equivalent in this case. This changes once gravity is coupled to the Standard Model, and the two formulations give different predictions. \textit{A priori}, there is no irrefutable reason to prefer one or the other theory. Therefore, it is important to investigate implications of the EC formulation of gravity.

An interesting property of EC gravity is that it allows for additional invariants of mass dimension not bigger than four. They arise due to the non-minimal coupling of gravity either to scalars or to fermions. In this work, we have generalized previous results by including all such contributions that are not present in metric gravity. They are displayed in \eqs \eqref{S_gr} and \eqref{S_f}. As a first step towards investigating their implications, we have derived an equivalent formulation of the theory in metric gravity. The resulting effective action \eqref{S_eff} represents the main result of our paper. 

The next step is to study how the various additional terms affect cosmology and experiment. In \cite{secondPaper}, we investigate the implications of the higher-order scalar self-interactions for Higgs inflation. This leads to scenarios that generalize the known models of metric \cite{Bezrukov:2007ep} and Palatini \cite{Bauer:2008zj} Higgs inflation. We find that inflation is both possible and consistent with observations in a broad range of parameters (\ref{params}). Furthermore, the spectral index $n_s$ is in most cases given by $n_s=1-2/N_\star$, where $N_\star$ is the number of e-foldings. In contrast, the tensor-to-scalar ratio $r$ varies in a wide range $\sim (10^{-10},1)$. We also discuss the robustness of inflationary predictions against scalar-fermion and fermion-fermion interactions present in the effective theory (\ref{S_eff}). This leads to upper bounds on the non-minimal fermion couplings $\alpha$ and $\beta$.

There is an intriguing consequence of the four-fermion interactions present in action~\eqref{S_eff}. Namely, they can mediate the production of feebly interacting (Dirac or Majorana) fermions right after inflation. If such fermions are singlets with respect to the Standard Model gauge group, they can play the role of dark matter. This possibility is explored in \cite{thirdPaper} where we show that the observed dark matter abundance can be generated by the four-fermion interactions in a wide range of fermion masses -- from a keV-scale up to $\sim 10^8~$GeV -- while respecting the bounds on the non-minimal fermion couplings $\alpha$ and $\beta$ coming from Higgs inflation \cite{secondPaper}. Moreover, our proposed production mechanism leads to a characteristic primordial momentum distribution of dark matter \cite{thirdPaper}. In the case of keV-scale warm dark matter, it can potentially be observable. 
 A well-motivated example of fermion dark matter produced this way is one of the right-handed neutrinos of the $\nu$MSM~\cite{Asaka:2005an,Asaka:2005pn}. Since the production mechanism by the four-fermion interaction is also operative for an absolutely stable sterile neutrino, X-ray constraints disappear and masses above the keV-scale are no longer excluded.\footnote{For reviews of sterile neutrino dark matter see \cite{Boyarsky:2009ix,Boyarsky:2018tvu}.}

Finally, the above remarks also apply to a no-scale scenario \cite{Shaposhnikov:2008xb,GarciaBellido:2011de,Bezrukov:2012hx,2004.00039}, in which both the Standard Model and gravity contain no dimensionful parameters at the classical level. Instead, one adds an additional scalar degree of freedom --- dilaton --- and a scale-invariant potential for the dilaton and the Higgs fields. In this setting, we have considered all terms that are specific to the EC formulation of gravity (\eqs \eqref{S_gr_d}, \eqref{S_s_d} as well as \eqref{S_f} as before) and then solved the theory for torsion to obtain an equivalent theory in metric gravity. How the previous result \eqref{S_eff} changes is displayed in \eq\eqref{S_eff_d}. It would also be interesting to extend the phenomenological studies to this scenario.

\section*{Acknowledgements}

The work was supported by ERC-AdG-2015~grant~694896 and by the Swiss National Science Foundation Excellence grant~200020B\underline{ }182864.

\paragraph{Note added.} While the present work was in preparation, a paper which also addresses the computation of the scalar-gravity sector of the theory (\ref{S_tot}) appeared \cite{Langvik:2020nrs}.

\bibliographystyle{JHEP}
\bibliography{EC}

\providecommand{\href}[2]{#2}\begingroup\raggedright\begin{thebibliography}{10}

\bibitem{Cartan:1922}
{\'E}.~Cartan, \emph{Sur une g{\'e}n{\'e}ralisation de la notion de courbure de
  riemann et les espaces {\`a} torsion}, {\emph{Comptes Rendus, Ac. Sc. Paris}
  {\bfseries 174} (1922) 593}.

\bibitem{Cartan:1923}
{\'E}.~Cartan, \emph{Sur les vari{\'e}t{\'e}s {\`a} connexion affine et la
  th{\'e}orie de la relativit{\'e} g{\'e}n{\'e}ralis{\'e}e (premi{\`e}re
  partie)},  in \emph{Annales scientifiques de l'{\'E}cole normale
  sup{\'e}rieure}, vol.~40, pp.~325--412, 1923.

\bibitem{Cartan:1924}
{\'E}.~Cartan, \emph{Sur les vari{\'e}t{\'e}s {\`a} connexion affine, et la
  th{\'e}orie de la relativit{\'e} g{\'e}n{\'e}ralis{\'e}e (premi{\`e}re
  partie)(suite)},  in \emph{Annales scientifiques de l'{\'E}cole Normale
  Sup{\'e}rieure}, vol.~41, pp.~1--25, 1924.

\bibitem{Cartan1925}
{\'E}.~Cartan, \emph{Sur les vari{\'e}t{\'e}s {\`a} connexion affine, et la
  th{\'e}orie de la relativit{\'e} g{\'e}n{\'e}ralis{\'e}e (deuxi{\`e}me
  partie)},  in \emph{Annales scientifiques de l'{\'E}cole normale
  sup{\'e}rieure}, vol.~42, pp.~17--88, 1925.

\bibitem{Einstein1928}
A.~Einstein, \emph{Riemanngeometrie mit aufrechterhaltung des begriffes des
  fern-parallelismus}, {\emph{Sitzungsber. Preuss. Akad. Wiss} (1928) 217}.

\bibitem{Einstein19282}
A.~Einstein, \emph{Neue m{\"o}glichkeit f{\"u}r eine einheitliche feldtheorie
  von gravitation und elektrizit{\"a}t}, {\emph{Sitzungsber. Preuss. Akad.
  Wiss} (1928) 224}.

\bibitem{Utiyama:1956sy}
R.~Utiyama, \emph{{Invariant theoretical interpretation of interaction}},
  \href{https://doi.org/10.1103/PhysRev.101.1597}{\emph{Phys. Rev.} {\bfseries
  101} (1956) 1597}.

\bibitem{sciama1962analogy}
D.~W. Sciama, \emph{On the analogy between charge and spin in general
  relativity}, {\emph{Recent developments in general relativity} (1962) 415}.

\bibitem{Kibble:1961ba}
T.~W.~B. Kibble, \emph{{Lorentz invariance and the gravitational field}},
  \href{https://doi.org/10.1063/1.1703702}{\emph{J. Math. Phys.} {\bfseries 2}
  (1961) 212}.

\bibitem{Hehl:1976kj}
F.~W. Hehl, P.~Von Der~Heyde, G.~D. Kerlick and J.~M. Nester, \emph{{General
  Relativity with Spin and Torsion: Foundations and Prospects}},
  \href{https://doi.org/10.1103/RevModPhys.48.393}{\emph{Rev. Mod. Phys.}
  {\bfseries 48} (1976) 393}.

\bibitem{Shapiro:2001rz}
I.~L. Shapiro, \emph{{Physical aspects of the space-time torsion}},
  \href{https://doi.org/10.1016/S0370-1573(01)00030-8}{\emph{Phys. Rept.}
  {\bfseries 357} (2002) 113}
  [\href{https://arxiv.org/abs/hep-th/0103093}{{\ttfamily hep-th/0103093}}].

\bibitem{Hehl:1978yt}
F.~Hehl, Y.~Ne'eman, J.~Nitsch and P.~Von~der Heyde, \emph{{Short Range
  Confining Component in a Quadratic Poincare Gauge Theory of Gravitation}},
  \href{https://doi.org/10.1016/0370-2693(78)90358-1}{\emph{Phys. Lett. B}
  {\bfseries 78} (1978) 102}.

\bibitem{Hayashi:1979wj}
K.~Hayashi and T.~Shirafuji, \emph{{Gravity from Poincare Gauge Theory of the
  Fundamental Particles. 1. Linear and Quadratic Lagrangians}},
  \href{https://doi.org/10.1143/PTP.64.866}{\emph{Prog. Theor. Phys.}
  {\bfseries 64} (1980) 866}.

\bibitem{Sezgin:1979zf}
E.~Sezgin and P.~van Nieuwenhuizen, \emph{{New Ghost Free Gravity Lagrangians
  with Propagating Torsion}},
  \href{https://doi.org/10.1103/PhysRevD.21.3269}{\emph{Phys. Rev. D}
  {\bfseries 21} (1980) 3269}.

\bibitem{Bregman:1973fv}
A.~Bregman, \emph{{Weyl transformations and poincare gauge invariance}},
  \href{https://doi.org/10.1143/PTP.49.667}{\emph{Prog. Theor. Phys.}
  {\bfseries 49} (1973) 667}.

\bibitem{Charap:1973fi}
J.~M. Charap and W.~Tait, \emph{{A gauge theory of the Weyl group}},
  \href{https://doi.org/10.1098/rspa.1974.0151}{\emph{Proc. Roy. Soc. Lond.}
  {\bfseries A340} (1974) 249}.

\bibitem{Kasuya:1975tx}
M.~Kasuya, \emph{{On the Gauge Theory in the Einstein-Cartan-Weyl Space-Time}},
  \href{https://doi.org/10.1007/BF02722810}{\emph{Nuovo Cim.} {\bfseries B28}
  (1975) 127}.

\bibitem{blagojevic2013gauge}
M.~Blagojevi{\'c} and F.~Hehl, \emph{Gauge Theories of Gravitation: A Reader
  with Commentaries}, Classification of Gauge Theories of Gravity. Imperial
  College Press, 2013.

\bibitem{Krasnov:2017epi}
K.~Krasnov and R.~Percacci, \emph{{Gravity and Unification: A review}},
  \href{https://doi.org/10.1088/1361-6382/aac58d}{\emph{Class. Quant. Grav.}
  {\bfseries 35} (2018) 143001}
  [\href{https://arxiv.org/abs/1712.03061}{{\ttfamily 1712.03061}}].

\bibitem{Ivanov:1981wn}
E.~A. Ivanov and J.~Niederle, \emph{{Gauge Formulation of Gravitation Theories.
  1. The Poincare, De Sitter and Conformal Cases}},
  \href{https://doi.org/10.1103/PhysRevD.25.976}{\emph{Phys. Rev.} {\bfseries
  D25} (1982) 976}.

\bibitem{Nair:2008yh}
V.~Nair, S.~Randjbar-Daemi and V.~Rubakov, \emph{{Massive Spin-2 fields of
  Geometric Origin in Curved Spacetimes}},
  \href{https://doi.org/10.1103/PhysRevD.80.104031}{\emph{Phys. Rev. D}
  {\bfseries 80} (2009) 104031}
  [\href{https://arxiv.org/abs/0811.3781}{{\ttfamily 0811.3781}}].

\bibitem{Nikiforova:2009qr}
V.~Nikiforova, S.~Randjbar-Daemi and V.~Rubakov, \emph{{Infrared Modified
  Gravity with Dynamical Torsion}},
  \href{https://doi.org/10.1103/PhysRevD.80.124050}{\emph{Phys. Rev. D}
  {\bfseries 80} (2009) 124050}
  [\href{https://arxiv.org/abs/0905.3732}{{\ttfamily 0905.3732}}].

\bibitem{Baekler:2010fr}
P.~Baekler, F.~W. Hehl and J.~M. Nester, \emph{{Poincare gauge theory of
  gravity: Friedman cosmology with even and odd parity modes. Analytic part}},
  \href{https://doi.org/10.1103/PhysRevD.83.024001}{\emph{Phys. Rev. D}
  {\bfseries 83} (2011) 024001}
  [\href{https://arxiv.org/abs/1009.5112}{{\ttfamily 1009.5112}}].

\bibitem{Karananas:2014pxa}
G.~K. Karananas, \emph{{The particle spectrum of parity-violating Poincaré
  gravitational theory}},
  \href{https://doi.org/10.1088/0264-9381/32/5/055012}{\emph{Class. Quant.
  Grav.} {\bfseries 32} (2015) 055012}
  [\href{https://arxiv.org/abs/1411.5613}{{\ttfamily 1411.5613}}].

\bibitem{Lasenby:2015dba}
A.~Lasenby and M.~Hobson, \emph{{Scale-invariant gauge theories of gravity:
  theoretical foundations}}, \href{https://doi.org/10.1063/1.4963143}{\emph{J.
  Math. Phys.} {\bfseries 57} (2016) 092505}
  [\href{https://arxiv.org/abs/1510.06699}{{\ttfamily 1510.06699}}].

\bibitem{York:1972sj}
J.~W. York, Jr., \emph{{Role of conformal three geometry in the dynamics of
  gravitation}}, \href{https://doi.org/10.1103/PhysRevLett.28.1082}{\emph{Phys.
  Rev. Lett.} {\bfseries 28} (1972) 1082}.

\bibitem{Gibbons:1976ue}
G.~W. Gibbons and S.~W. Hawking, \emph{{Action Integrals and Partition
  Functions in Quantum Gravity}},
  \href{https://doi.org/10.1103/PhysRevD.15.2752}{\emph{Phys. Rev.} {\bfseries
  D15} (1977) 2752}.

\bibitem{Dadhich:2010xa}
N.~Dadhich and J.~M. Pons, \emph{{On the equivalence of the Einstein-Hilbert
  and the Einstein-Palatini formulations of general relativity for an arbitrary
  connection}}, \href{https://doi.org/10.1007/s10714-012-1393-9}{\emph{Gen.
  Rel. Grav.} {\bfseries 44} (2012) 2337}
  [\href{https://arxiv.org/abs/1010.0869}{{\ttfamily 1010.0869}}].

\bibitem{Palatini1919}
A.~Palatini, \emph{Deduzione invariantiva delle equazioni gravitazionali dal
  principio di hamilton},
  \href{https://doi.org/10.1007/BF03014670}{\emph{Rendiconti del Circolo
  Matematico di Palermo} {\bfseries 43} (1919) 203}.

\bibitem{Einstein1925}
A.~Einstein, \emph{{Einheitliche Feldtheorie von Gravitation und
  Elektrizit\"at}}, {\emph{Sitzungsber. Preuss. Akad. Wiss} {\bfseries 414}
  (1925) }.

\bibitem{Ferraris1981}
M.~Ferraris, M.~Francaviglia and C.~Reina, \emph{Variational formulation of
  general relativity from 1915 to 1925 ``palatini's method''discovered by
  einstein in 1925}, \href{https://doi.org/10.1007/BF00756060}{\emph{General
  Relativity and Gravitation} {\bfseries 14} (1982) 243}.

\bibitem{osti_4843429}
V.~I. Rodichev, \emph{Twisted space and nonlinear field equations},
  {\emph{Zhur. Eksptl'. i Teoret. Fiz.} {\bfseries 40} (1961) }.

\bibitem{1810.05536}
V.-M. Enckell, K.~Enqvist, S.~Rasanen and L.-P. Wahlman, \emph{{Inflation with
  $R^2$ term in the Palatini formalism}},
  \href{https://doi.org/10.1088/1475-7516/2019/02/022}{\emph{JCAP} {\bfseries
  1902} (2019) 022} [\href{https://arxiv.org/abs/1810.05536}{{\ttfamily
  1810.05536}}].

\bibitem{1810.10418}
I.~Antoniadis, A.~Karam, A.~Lykkas and K.~Tamvakis, \emph{{Palatini inflation
  in models with an $R^2$ term}},
  \href{https://doi.org/10.1088/1475-7516/2018/11/028}{\emph{JCAP} {\bfseries
  1811} (2018) 028} [\href{https://arxiv.org/abs/1810.10418}{{\ttfamily
  1810.10418}}].

\bibitem{1901.01794}
T.~Tenkanen, \emph{{Minimal Higgs inflation with an $R^2$ term in Palatini
  gravity}}, \href{https://doi.org/10.1103/PhysRevD.99.063528}{\emph{Phys.
  Rev.} {\bfseries D99} (2019) 063528}
  [\href{https://arxiv.org/abs/1901.01794}{{\ttfamily 1901.01794}}].

\bibitem{1902.07876}
A.~Edery and Y.~Nakayama, \emph{{Palatini formulation of pure $R^2$ gravity
  yields Einstein gravity with no massless scalar}},
  \href{https://doi.org/10.1103/PhysRevD.99.124018}{\emph{Phys. Rev.}
  {\bfseries D99} (2019) 124018}
  [\href{https://arxiv.org/abs/1902.07876}{{\ttfamily 1902.07876}}].

\bibitem{1911.11513}
I.~D. Gialamas and A.~B. Lahanas, \emph{{Reheating in $R^2$ Palatini
  inflationary models}},
  \href{https://doi.org/10.1103/PhysRevD.101.084007}{\emph{Phys. Rev.}
  {\bfseries D101} (2020) 084007}
  [\href{https://arxiv.org/abs/1911.11513}{{\ttfamily 1911.11513}}].

\bibitem{1912.12757}
I.~Antoniadis, A.~Karam, A.~Lykkas, T.~Pappas and K.~Tamvakis,
  \emph{{Single-field inflation in models with an $R^2$ term}},  in \emph{{19th
  Hellenic School and Workshops on Elementary Particle Physics and Gravity
  (CORFU2019) Corfu, Greece, August 31-September 25, 2019}}, 2019,
  \href{https://arxiv.org/abs/1912.12757}{{\ttfamily 1912.12757}}.

\bibitem{2002.08324}
A.~Lloyd-Stubbs and J.~McDonald, \emph{{Sub-Planckian $\phi^{2}$ Inflation in
  the Palatini Formulation of Gravity with an $R^2$ term}},
  \href{https://arxiv.org/abs/2002.08324}{{\ttfamily 2002.08324}}.

\bibitem{Hojman:1980kv}
R.~Hojman, C.~Mukku and W.~A. Sayed, \emph{{Parity violation in metric torsion
  theories of gravitation}},
  \href{https://doi.org/10.1103/PhysRevD.22.1915}{\emph{Phys. Rev.} {\bfseries
  D22} (1980) 1915}.

\bibitem{Nelson:1980ph}
P.~C. Nelson, \emph{{Gravity With Propagating Pseudoscalar Torsion}},
  \href{https://doi.org/10.1016/0375-9601(80)90348-5}{\emph{Phys. Lett.}
  {\bfseries A79} (1980) 285}.

\bibitem{Castellani:1991et}
L.~Castellani, R.~D'Auria and P.~Fre, \emph{{Supergravity and superstrings: A
  Geometric perspective. Vol. 1: Mathematical foundations}}. 1991.

\bibitem{Holst:1995pc}
S.~Holst, \emph{{Barbero's Hamiltonian derived from a generalized
  Hilbert-Palatini action}},
  \href{https://doi.org/10.1103/PhysRevD.53.5966}{\emph{Phys. Rev.} {\bfseries
  D53} (1996) 5966} [\href{https://arxiv.org/abs/gr-qc/9511026}{{\ttfamily
  gr-qc/9511026}}].

\bibitem{Nieh:1981ww}
H.~T. Nieh and M.~L. Yan, \emph{{An Identity in Riemann-cartan Geometry}},
  \href{https://doi.org/10.1063/1.525379}{\emph{J. Math. Phys.} {\bfseries 23}
  (1982) 373}.

\bibitem{hep-th/0507253}
L.~Freidel, D.~Minic and T.~Takeuchi, \emph{{Quantum gravity, torsion, parity
  violation and all that}},
  \href{https://doi.org/10.1103/PhysRevD.72.104002}{\emph{Phys. Rev.}
  {\bfseries D72} (2005) 104002}
  [\href{https://arxiv.org/abs/hep-th/0507253}{{\ttfamily hep-th/0507253}}].

\bibitem{Alexandrov:2008iy}
S.~Alexandrov, \emph{{Immirzi parameter and fermions with non-minimal
  coupling}},
  \href{https://doi.org/10.1088/0264-9381/25/14/145012}{\emph{Class. Quant.
  Grav.} {\bfseries 25} (2008) 145012}
  [\href{https://arxiv.org/abs/0802.1221}{{\ttfamily 0802.1221}}].

\bibitem{1104.2432}
D.~Diakonov, A.~G. Tumanov and A.~A. Vladimirov, \emph{{Low-energy General
  Relativity with torsion: A Systematic derivative expansion}},
  \href{https://doi.org/10.1103/PhysRevD.84.124042}{\emph{Phys. Rev.}
  {\bfseries D84} (2011) 124042}
  [\href{https://arxiv.org/abs/1104.2432}{{\ttfamily 1104.2432}}].

\bibitem{1212.0585}
J.~Magueijo, T.~G. Zlosnik and T.~W.~B. Kibble, \emph{{Cosmology with a spin}},
  \href{https://doi.org/10.1103/PhysRevD.87.063504}{\emph{Phys. Rev.}
  {\bfseries D87} (2013) 063504}
  [\href{https://arxiv.org/abs/1212.0585}{{\ttfamily 1212.0585}}].

\bibitem{gr-qc/0505081}
A.~Perez and C.~Rovelli, \emph{{Physical effects of the Immirzi parameter}},
  \href{https://doi.org/10.1103/PhysRevD.73.044013}{\emph{Phys. Rev.}
  {\bfseries D73} (2006) 044013}
  [\href{https://arxiv.org/abs/gr-qc/0505081}{{\ttfamily gr-qc/0505081}}].

\bibitem{0807.2652}
V.~Taveras and N.~Yunes, \emph{{The Barbero-Immirzi Parameter as a Scalar
  Field: K-Inflation from Loop Quantum Gravity?}},
  \href{https://doi.org/10.1103/PhysRevD.78.064070}{\emph{Phys. Rev.}
  {\bfseries D78} (2008) 064070}
  [\href{https://arxiv.org/abs/0807.2652}{{\ttfamily 0807.2652}}].

\bibitem{0811.1998}
A.~Torres-Gomez and K.~Krasnov, \emph{{Remarks on Barbero-Immirzi parameter as
  a field}}, \href{https://doi.org/10.1103/PhysRevD.79.104014}{\emph{Phys.
  Rev.} {\bfseries D79} (2009) 104014}
  [\href{https://arxiv.org/abs/0811.1998}{{\ttfamily 0811.1998}}].

\bibitem{0902.0957}
G.~Calcagni and S.~Mercuri, \emph{{The Barbero-Immirzi field in canonical
  formalism of pure gravity}},
  \href{https://doi.org/10.1103/PhysRevD.79.084004}{\emph{Phys. Rev.}
  {\bfseries D79} (2009) 084004}
  [\href{https://arxiv.org/abs/0902.0957}{{\ttfamily 0902.0957}}].

\bibitem{0902.2764}
S.~Mercuri, \emph{{Peccei-Quinn mechanism in gravity and the nature of the
  Barbero-Immirzi parameter}},
  \href{https://doi.org/10.1103/PhysRevLett.103.081302}{\emph{Phys. Rev. Lett.}
  {\bfseries 103} (2009) 081302}
  [\href{https://arxiv.org/abs/0902.2764}{{\ttfamily 0902.2764}}].

\bibitem{Ashtekar:1986yd}
A.~Ashtekar, \emph{{New Variables for Classical and Quantum Gravity}},
  \href{https://doi.org/10.1103/PhysRevLett.57.2244}{\emph{Phys. Rev. Lett.}
  {\bfseries 57} (1986) 2244}.

\bibitem{Barbero:1994ap}
J.~Barbero~G., \emph{{Real Ashtekar variables for Lorentzian signature space
  times}}, \href{https://doi.org/10.1103/PhysRevD.51.5507}{\emph{Phys. Rev. D}
  {\bfseries 51} (1995) 5507}
  [\href{https://arxiv.org/abs/gr-qc/9410014}{{\ttfamily gr-qc/9410014}}].

\bibitem{Thiemann:2007zz}
T.~Thiemann, \emph{{Modern canonical quantum general relativity}},
  \href{https://arxiv.org/abs/gr-qc/0110034}{{\ttfamily gr-qc/0110034}}.

\bibitem{1201.4226}
I.~B. Khriplovich, \emph{{Gravitational four-fermion interaction on the Planck
  scale}}, \href{https://doi.org/10.1016/j.physletb.2012.01.072}{\emph{Phys.
  Lett.} {\bfseries B709} (2012) 111}
  [\href{https://arxiv.org/abs/1201.4226}{{\ttfamily 1201.4226}}].

\bibitem{1201.5423}
G.~de~Berredo-Peixoto, L.~Freidel, I.~L. Shapiro and C.~A. de~Souza,
  \emph{{Dirac fields, torsion and Barbero-Immirzi parameter in Cosmology}},
  \href{https://doi.org/10.1088/1475-7516/2012/06/017}{\emph{JCAP} {\bfseries
  1206} (2012) 017} [\href{https://arxiv.org/abs/1201.5423}{{\ttfamily
  1201.5423}}].

\bibitem{Khriplovich:2013tqa}
I.~B. Khriplovich and A.~S. Rudenko, \emph{{Gravitational four-fermion
  interaction and dynamics of the early Universe}},
  \href{https://doi.org/10.1007/JHEP11(2013)174}{\emph{JHEP} {\bfseries 11}
  (2013) 174} [\href{https://arxiv.org/abs/1303.1348}{{\ttfamily 1303.1348}}].

\bibitem{secondPaper}
M.~Shaposhnikov, A.~Shkerin, I.~Timiryasov and S.~Zell, \emph{{Higgs inflation
  in Einstein-Cartan gravity}},
  \href{https://doi.org/10.1088/1475-7516/2021/02/008}{\emph{JCAP} {\bfseries
  02} (2021) 008} [\href{https://arxiv.org/abs/2007.14978}{{\ttfamily
  2007.14978}}].

\bibitem{Bezrukov:2007ep}
F.~L. Bezrukov and M.~Shaposhnikov, \emph{{The Standard Model Higgs boson as
  the inflaton}},
  \href{https://doi.org/10.1016/j.physletb.2007.11.072}{\emph{Phys. Lett.}
  {\bfseries B659} (2008) 703}
  [\href{https://arxiv.org/abs/0710.3755}{{\ttfamily 0710.3755}}].

\bibitem{Bauer:2008zj}
F.~Bauer and D.~A. Demir, \emph{{Inflation with Non-Minimal Coupling: Metric
  versus Palatini Formulations}},
  \href{https://doi.org/10.1016/j.physletb.2008.06.014}{\emph{Phys. Lett.}
  {\bfseries B665} (2008) 222}
  [\href{https://arxiv.org/abs/0803.2664}{{\ttfamily 0803.2664}}].

\bibitem{Shaposhnikov:2020fdv}
M.~Shaposhnikov, A.~Shkerin and S.~Zell, \emph{{Quantum Effects in Palatini
  Higgs Inflation}},
  \href{https://doi.org/10.1088/1475-7516/2020/07/064}{\emph{JCAP} {\bfseries
  07} (2020) 064} [\href{https://arxiv.org/abs/2002.07105}{{\ttfamily
  2002.07105}}].

\bibitem{Shaposhnikov:2020geh}
M.~Shaposhnikov, A.~Shkerin and S.~Zell, \emph{{Standard Model Meets Gravity:
  Electroweak Symmetry Breaking and Inflation}},
  \href{https://doi.org/10.1103/PhysRevD.103.033006}{\emph{Phys. Rev. D}
  {\bfseries 103} (2021) 033006}
  [\href{https://arxiv.org/abs/2001.09088}{{\ttfamily 2001.09088}}].

\bibitem{Shaposhnikov:2018xkv}
M.~Shaposhnikov and A.~Shkerin, \emph{{Conformal symmetry: towards the link
  between the Fermi and the Planck scales}},
  \href{https://doi.org/10.1016/j.physletb.2018.06.068}{\emph{Phys. Lett. B}
  {\bfseries 783} (2018) 253}
  [\href{https://arxiv.org/abs/1803.08907}{{\ttfamily 1803.08907}}].

\bibitem{Shaposhnikov:2018jag}
M.~Shaposhnikov and A.~Shkerin, \emph{{Gravity, Scale Invariance and the
  Hierarchy Problem}},
  \href{https://doi.org/10.1007/JHEP10(2018)024}{\emph{JHEP} {\bfseries 10}
  (2018) 024} [\href{https://arxiv.org/abs/1804.06376}{{\ttfamily
  1804.06376}}].

\bibitem{Shkerin:2019mmu}
A.~Shkerin, \emph{{Dilaton-assisted generation of the Fermi scale from the
  Planck scale}}, \href{https://doi.org/10.1103/PhysRevD.99.115018}{\emph{Phys.
  Rev. D} {\bfseries 99} (2019) 115018}
  [\href{https://arxiv.org/abs/1903.11317}{{\ttfamily 1903.11317}}].

\bibitem{Karananas:2020qkp}
G.~K. Karananas, M.~Michel and J.~Rubio, \emph{{One residue to rule them all:
  Electroweak symmetry breaking, inflation and field-space geometry}},
  \href{https://doi.org/10.1016/j.physletb.2020.135876}{\emph{Phys. Lett. B}
  {\bfseries 811} (2020) 135876}
  [\href{https://arxiv.org/abs/2006.11290}{{\ttfamily 2006.11290}}].

\bibitem{thirdPaper}
M.~Shaposhnikov, A.~Shkerin, I.~Timiryasov and S.~Zell, \emph{{Einstein-Cartan
  Portal to Dark Matter}},
  \href{https://doi.org/10.1103/PhysRevLett.126.161301}{\emph{Phys. Rev. Lett.}
  {\bfseries 126} (2021) 161301}
  [\href{https://arxiv.org/abs/2008.11686}{{\ttfamily 2008.11686}}].

\bibitem{Asaka:2005an}
T.~Asaka, S.~Blanchet and M.~Shaposhnikov, \emph{{The nuMSM, dark matter and
  neutrino masses}},
  \href{https://doi.org/10.1016/j.physletb.2005.09.070}{\emph{Phys. Lett. B}
  {\bfseries 631} (2005) 151}
  [\href{https://arxiv.org/abs/hep-ph/0503065}{{\ttfamily hep-ph/0503065}}].

\bibitem{Asaka:2005pn}
T.~Asaka and M.~Shaposhnikov, \emph{{The $\nu$MSM, dark matter and baryon
  asymmetry of the universe}},
  \href{https://doi.org/10.1016/j.physletb.2005.06.020}{\emph{Phys. Lett. B}
  {\bfseries 620} (2005) 17}
  [\href{https://arxiv.org/abs/hep-ph/0505013}{{\ttfamily hep-ph/0505013}}].

\bibitem{Wetterich:1987fm}
C.~Wetterich, \emph{{Cosmology and the Fate of Dilatation Symmetry}},
  \href{https://doi.org/10.1016/0550-3213(88)90193-9}{\emph{Nucl. Phys. B}
  {\bfseries 302} (1988) 668}
  [\href{https://arxiv.org/abs/1711.03844}{{\ttfamily 1711.03844}}].

\bibitem{Shaposhnikov:2008xb}
M.~Shaposhnikov and D.~Zenhausern, \emph{{Scale invariance, unimodular gravity
  and dark energy}},
  \href{https://doi.org/10.1016/j.physletb.2008.11.054}{\emph{Phys. Lett.}
  {\bfseries B671} (2009) 187}
  [\href{https://arxiv.org/abs/0809.3395}{{\ttfamily 0809.3395}}].

\bibitem{GarciaBellido:2011de}
J.~Garcia-Bellido, J.~Rubio, M.~Shaposhnikov and D.~Zenhausern,
  \emph{{Higgs-Dilaton Cosmology: From the Early to the Late Universe}},
  \href{https://doi.org/10.1103/PhysRevD.84.123504}{\emph{Phys. Rev.}
  {\bfseries D84} (2011) 123504}
  [\href{https://arxiv.org/abs/1107.2163}{{\ttfamily 1107.2163}}].

\bibitem{Bezrukov:2012hx}
F.~Bezrukov, G.~K. Karananas, J.~Rubio and M.~Shaposhnikov,
  \emph{{Higgs-Dilaton Cosmology: an effective field theory approach}},
  \href{https://doi.org/10.1103/PhysRevD.87.096001}{\emph{Phys. Rev.}
  {\bfseries D87} (2013) 096001}
  [\href{https://arxiv.org/abs/1212.4148}{{\ttfamily 1212.4148}}].

\bibitem{Ferreira:2016wem}
P.~G. Ferreira, C.~T. Hill and G.~G. Ross, \emph{{Weyl Current, Scale-Invariant
  Inflation and Planck Scale Generation}},
  \href{https://doi.org/10.1103/PhysRevD.95.043507}{\emph{Phys. Rev. D}
  {\bfseries 95} (2017) 043507}
  [\href{https://arxiv.org/abs/1610.09243}{{\ttfamily 1610.09243}}].

\bibitem{Ferreira:2016kxi}
P.~G. Ferreira, C.~T. Hill and G.~G. Ross, \emph{{No fifth force in a scale
  invariant universe}},
  \href{https://doi.org/10.1103/PhysRevD.95.064038}{\emph{Phys. Rev. D}
  {\bfseries 95} (2017) 064038}
  [\href{https://arxiv.org/abs/1612.03157}{{\ttfamily 1612.03157}}].

\bibitem{Ferreira:2018itt}
P.~G. Ferreira, C.~T. Hill and G.~G. Ross, \emph{{Inertial Spontaneous Symmetry
  Breaking and Quantum Scale Invariance}},
  \href{https://doi.org/10.1103/PhysRevD.98.116012}{\emph{Phys. Rev. D}
  {\bfseries 98} (2018) 116012}
  [\href{https://arxiv.org/abs/1801.07676}{{\ttfamily 1801.07676}}].

\bibitem{Wetterich:2020cxq}
C.~Wetterich, \emph{{Fundamental Scale Invariance}},
  \href{https://arxiv.org/abs/2007.08805}{{\ttfamily 2007.08805}}.

\bibitem{2004.00039}
J.~Rubio, \emph{{Scale symmetry, the Higgs and the Cosmos}},  in \emph{{19th
  Hellenic School and Workshops on Elementary Particle Physics and Gravity
  (CORFU2019) Corfu, Greece, August 31-September 25, 2019}}, 2020,
  \href{https://arxiv.org/abs/2004.00039}{{\ttfamily 2004.00039}}.

\bibitem{Immirzi:1996dr}
G.~Immirzi, \emph{{Quantum gravity and Regge calculus}},
  \href{https://doi.org/10.1016/S0920-5632(97)00354-X}{\emph{Nucl. Phys. B
  Proc. Suppl.} {\bfseries 57} (1997) 65}
  [\href{https://arxiv.org/abs/gr-qc/9701052}{{\ttfamily gr-qc/9701052}}].

\bibitem{Immirzi:1996di}
G.~Immirzi, \emph{{Real and complex connections for canonical gravity}},
  \href{https://doi.org/10.1088/0264-9381/14/10/002}{\emph{Class. Quant. Grav.}
  {\bfseries 14} (1997) L177}
  [\href{https://arxiv.org/abs/gr-qc/9612030}{{\ttfamily gr-qc/9612030}}].

\bibitem{Khriplovich:2001qv}
I.~B. Khriplovich and R.~V. Korkin, \emph{{How is the maximum entropy of a
  quantized surface related to its area?}},
  \href{https://doi.org/10.1134/1.1499895}{\emph{J. Exp. Theor. Phys.}
  {\bfseries 95} (2002) 1}
  [\href{https://arxiv.org/abs/gr-qc/0112074}{{\ttfamily gr-qc/0112074}}].

\bibitem{Boyarsky:2009ix}
A.~Boyarsky, O.~Ruchayskiy and M.~Shaposhnikov, \emph{{The Role of sterile
  neutrinos in cosmology and astrophysics}},
  \href{https://doi.org/10.1146/annurev.nucl.010909.083654}{\emph{Ann. Rev.
  Nucl. Part. Sci.} {\bfseries 59} (2009) 191}
  [\href{https://arxiv.org/abs/0901.0011}{{\ttfamily 0901.0011}}].

\bibitem{Boyarsky:2018tvu}
A.~Boyarsky, M.~Drewes, T.~Lasserre, S.~Mertens and O.~Ruchayskiy,
  \emph{{Sterile neutrino Dark Matter}},
  \href{https://doi.org/10.1016/j.ppnp.2018.07.004}{\emph{Prog. Part. Nucl.
  Phys.} {\bfseries 104} (2019) 1}
  [\href{https://arxiv.org/abs/1807.07938}{{\ttfamily 1807.07938}}].

\bibitem{Langvik:2020nrs}
M.~L\r{a}ngvik, J.-M. Ojanper\"a, S.~Raatikainen and S.~R\"as\"anen,
  \emph{{Higgs inflation with the Holst and the Nieh\textendash{}Yan term}},
  \href{https://doi.org/10.1103/PhysRevD.103.083514}{\emph{Phys. Rev. D}
  {\bfseries 103} (2021) 083514}
  [\href{https://arxiv.org/abs/2007.12595}{{\ttfamily 2007.12595}}].

\end{thebibliography}\endgroup

\end{document}